\pdfoutput=1 % only if pdf/png/jpg images are used
\documentclass[a4paper,10pt]{article}
\usepackage{geometry}
\usepackage{graphics}  
\usepackage{graphicx}
\usepackage{subfigure}
\usepackage{amsmath}
\usepackage[load-configurations=abbreviations,tight-spacing=true,separate-uncertainty,bracket-numbers = false]{siunitx} 
\usepackage{nicefrac}
\usepackage[english]{babel}
\usepackage{lineno}
\usepackage{epstopdf}
\usepackage{stfloats}
\usepackage{verbatim}
\usepackage{url}
\usepackage{xcolor}
\usepackage{url}
\usepackage{setspace}
\usepackage{indentfirst}
\usepackage{txfonts}
\usepackage{float}
\usepackage{graphicx}

\newcommand*{\notFOREPJ}{}%

\setpagewiselinenumbers
\modulolinenumbers[5]

\ifdefined\notFOREPJ
\else
\fi

\makeatletter
\renewcommand{\maketitle}{\bgroup\setlength{\parindent}{0pt}
\begin{flushleft}
  \vspace*{10mm}
  \textbf{\huge\sffamily\@title}
  \vspace{5mm}
   
  \large \@author
\end{flushleft}\egroup
}
\def\@xfootnote[#1]{%
  \protected@xdef\@thefnmark{#1}%
  \@footnotemark\@footnotetext}
\makeatother

\begin{document}
\ifdefined\notFOREPJ
%\linenumbers
\else
\fi

\title{Development and simulations of Enhanced Lateral Drift Sensors}
\author{
A.~Velyka${}^{\textrm{a,}}$\footnote[*]{Corresponding author: anastasiia.velyka@desy.de}, %
H.~Jansen${}^{\textrm{a}}$
\\
\vspace{3mm}
${}^{\textrm{a}}$ Deutsches Elektronen-Synchrotron DESY, Hamburg, Germany\\
\vspace{3mm}
}
\maketitle

\begin{abstract}
We present the concept of a new type of silicon tracking sensor called Enhanced Lateral Drift (ELAD) sensor. 
In ELAD sensors the spatial resolution of the impact position of ionising particles is improved by a dedicated charge sharing mechanism,
 which is achieved by a non-homogeneous electric field in the lateral direction in the sensor bulk. 
The non-homogeneous electric field is created by buried doping implants with a higher concentration with respect to the background concentration of the bulk. 
The resulting position-dependent charge sharing allows for an improved interpolation of the impact position. 

TCAD-based electric field simulations for 2D and 3D geometries as well as transient simulations with a traversing particle for the 2D geometry have been carried out.
The electric field profiles have further been optimised for position resolution.
The simulations show a strong dependence of the charge sharing mechanism on the buried implant concentration.
Optimal values for the buried implant concentration allow for nearly linear charge sharing between two readout electrodes as a function of the impact position.
Additionally, the foreseen production technique combining silicon epitaxy and ion beam implantation is outlined.\\

\end{abstract}

\section{Introduction}
\label{sec:intro}

Experiments at possible future colliders like CLIC\footnote{Compact Linear Collider}~\cite{cdr} and ILC\footnote{International Linear Collider}~\cite{ILCcdr} aim, among others,
 for a precise measurement of Higgs decays to pairs of b-quarks, c-quarks and gluons and efficient identification of top quarks in the decay t$\mathrm{\rightarrow}$Wb.
To this end, the experiments require light-weight detectors with a single-point resolution of a few micrometers. 
E.g., the tracker and vertex detectors for CLIC require a single point resolution of better than 7~$\muup$m and 3~$\muup$m, respectively, in the transverse plane
 to meet the requirements on the track-momentum and impact parameter resolution at a total silicon detector thickness of less than 100~$\muup$m~\cite{det}.
Several options for tracking and vertex sensor technology are considered for CLIC, among them are monolithic and hybrid detectors~\cite{clicyellowreport}.

This work focuses on a new type of a hybrid detector aiming to meet the requirements of the CLIC vertex detector.
A common approach to achieve an improved position resolution in silicon detectors is to decrease the readout cell size.
This leads to an increased number of readout channels and less area for logic per channel on the readout chip.
Additionally, the miniaturisation of interconnection techniques might present limits. 

The position resolution can also be improved by means of charge sharing, i.e.\ by collecting the charge on more than one readout electrode.
Geometric charge sharing is used e.g.\ in the CMS experiment by tilting the sensors in the Endcap~\cite{simon}. 
Lorentz drift induced charge sharing by means of a magnetic field is realised in the barrel section of the experiment~\cite{cmstdr}.
However, these two methods do not provide sufficient charge sharing in thin ($\leq 100\,\muup$m) sensors~\cite{clicyellowreport},
 as either unrealistically strong magnetic fields or huge sensor tilts would be necessary, the latter in turn increasing the material budget.

The hybrid detector proposed herein, a so-called Enhanced Lateral Drift (ELAD) sensor in combination with a readout ASIC,
 introduces a lateral electric field component in the sensor bulk yielding a lateral charge drift to achieve a position resolution of a few micrometre. 
This lateral component of the electric field is created by regions deep within the sensor bulk featuring a higher doping concentration with respect to the background doping concentration.
This diminishes the need to downsize the sensor pitch while avoiding sensor tilt or magnetic fields. 

In Sec.~\ref{sec:con}, we detail the concept of the ELAD sensor.
The SYNOPSYS TCAD simulations and the analysis of charge collection distribution on neighbouring readout electrodes are discussed in Sec.~\ref{sec:sim}.
Finally, the production technique is outlined in Sec.~\ref{sec:pr}.

\section{Concept of the ELAD sensor}
\label{sec:con}
Charged particles traversing a silicon sensor create free electron-hole pairs, which drift inside the sensor following the electric field lines.
Due to insufficient diffusion in thin, depleted, standard planar sensors the created charge is collected almost entirely by the nearest electrode and only a small fraction of it by the neighbouring electrode.
Therefore, in this particular case, an improved position resolution is solely achieved by downsizing the strip/pixel pitch.

\begin{figure}[t]
  \centering
  \includegraphics[height=4.1cm]{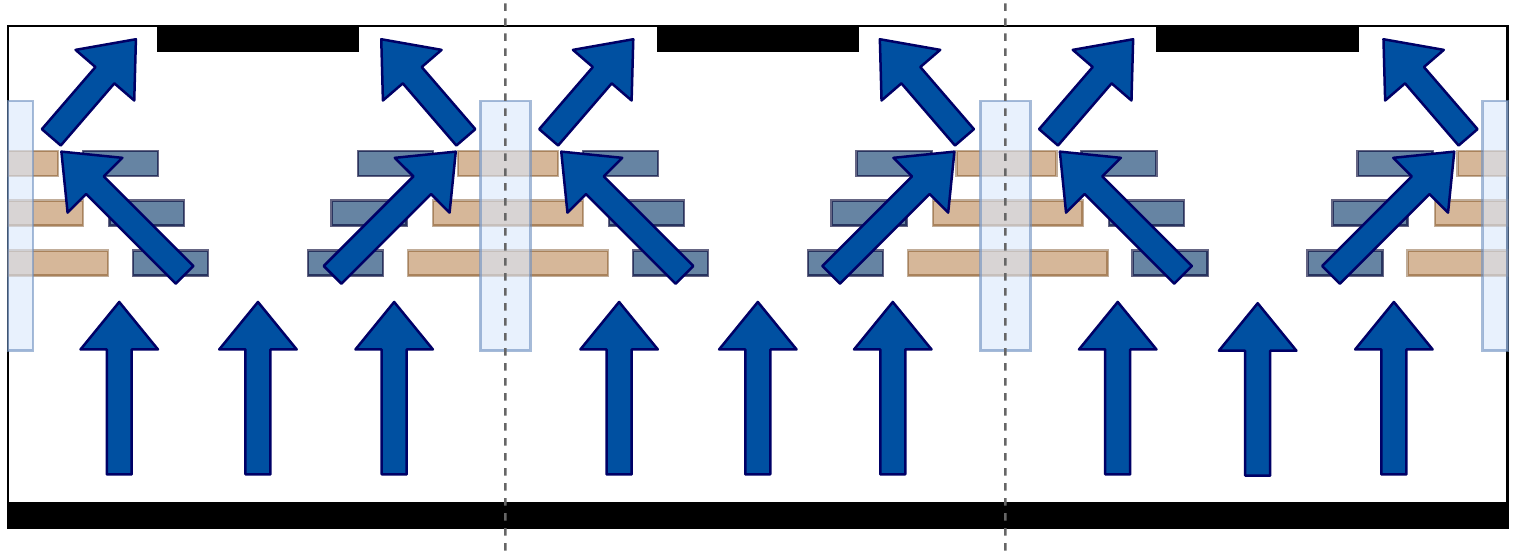}\put(-269, 115){RO}\put(-225, 125){\footnotesize unit cell}\put(-227.5, 117){\footnotesize boundary}\put(-165, 115){RO}\put(-120, 125){\footnotesize unit cell}\put(-122.5, 117){\footnotesize boundary}\put(-61, 115){RO}\put(-215, 50){\rotatebox{90}{\small diffusion}}\put(-110, 50){\rotatebox{90}{\small diffusion}}
  \caption{
A sketch of the charge sharing mechanism in ELAD sensors is shown. 
Arrows indicate the direction of the field strength. 
}
  \label{fig:concept}
\end{figure}
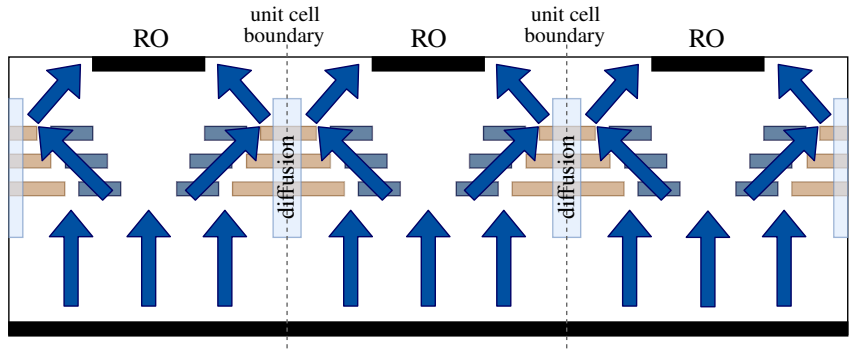

The concept of the ELAD sensor is based on a dedicated charge sharing mechanism independent of sensor tilt and magnetic fields.
Local modifications to the electric field yield a position-dependent charge collection at two electrodes~\cite{hj}.
To this end, the electric field profile is altered by buried implants with higher doping concentration with respect to the background doping concentration of the bulk. 
The buried implants are arranged in layers forming a trapezoidal area changing the electric field lines in such a way
 that the charge carriers, e.g.\ holes in an n-type sensor, drift towards the centre between two electrodes to possibly diffuse into the next unit cell, see Fig.~\ref{fig:concept}.
This scheme results in a partial collection at two electrodes.
An optimal position resolution is achieved if the distribution of the collected charge between two electrodes is a linear function of the particle's impact position, 
 i.e.\ the entirety of the charge is collected at one electrode if the impact position is in the centre of the readout electrode,
 the charge is equally shared between two electrodes if the impact position is half-way between two electrodes
 and in between the shared fraction of the charge increases linearly with the distance to the electrode's centre.
Our realisation of an ELAD sensor contains two types of buried implants: p-implants and n-implants, effectively reducing the impact of the additional stationary charges on $N\mathrm{_{eff}}$~\cite{elad}. 
With such a balancing of p- and n-implants, a similar depletion voltage as in a standard planar sensor with an epitaxial layer of the same background concentration and thickness is achieved.
The utilisation of two types of buried implants also creates a stronger electric field in the lateral direction, thereby improving the charge-sharing behaviour.

\section{TCAD simulations}
\label{sec:sim}

\begin{figure}[t!]
  \centering
  \includegraphics[trim=0.5cm 3.0cm 0.5cm 2.1cm, width = 0.18\textwidth, clip]{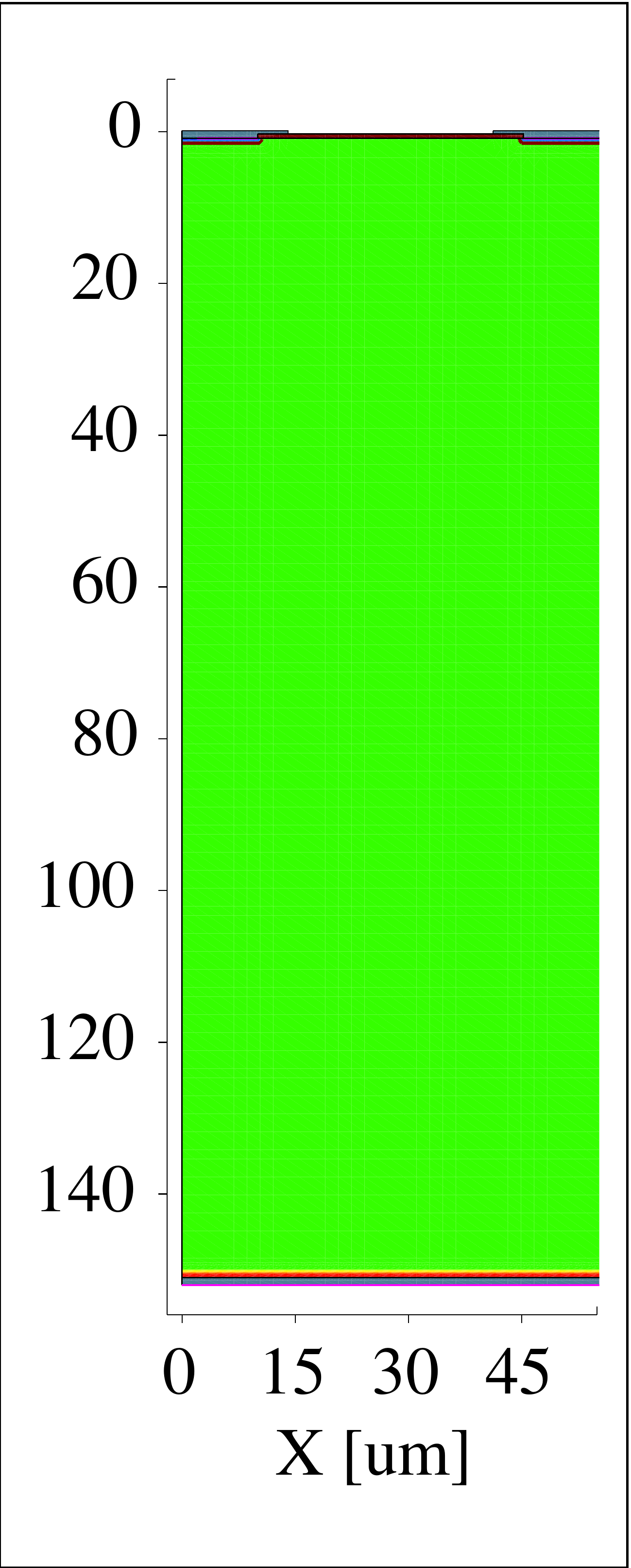}\put(-80,-10){a)}\put(-56, 175){$\mathrm{RO_{l}}$}\put(-17, 175){$\mathrm{RO_{r}}$}\put(-45,-10){\small X [$\muup$m]}\put(-88,100){\rotatebox{90}{\small Y [$\muup$m]}}
    \includegraphics[width = 0.025\textwidth ]{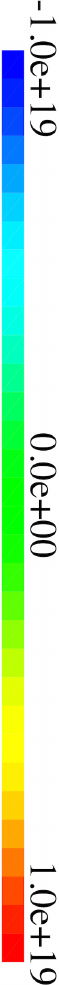}\put(0, 145){\rotatebox{-90}{\scriptsize Doping Concentration [$\mathrm{cm^{-3}}$]}}\hfill
  \includegraphics[trim=0.5cm 3.0cm 0.5cm 2.1cm, width = 0.18\textwidth, clip]{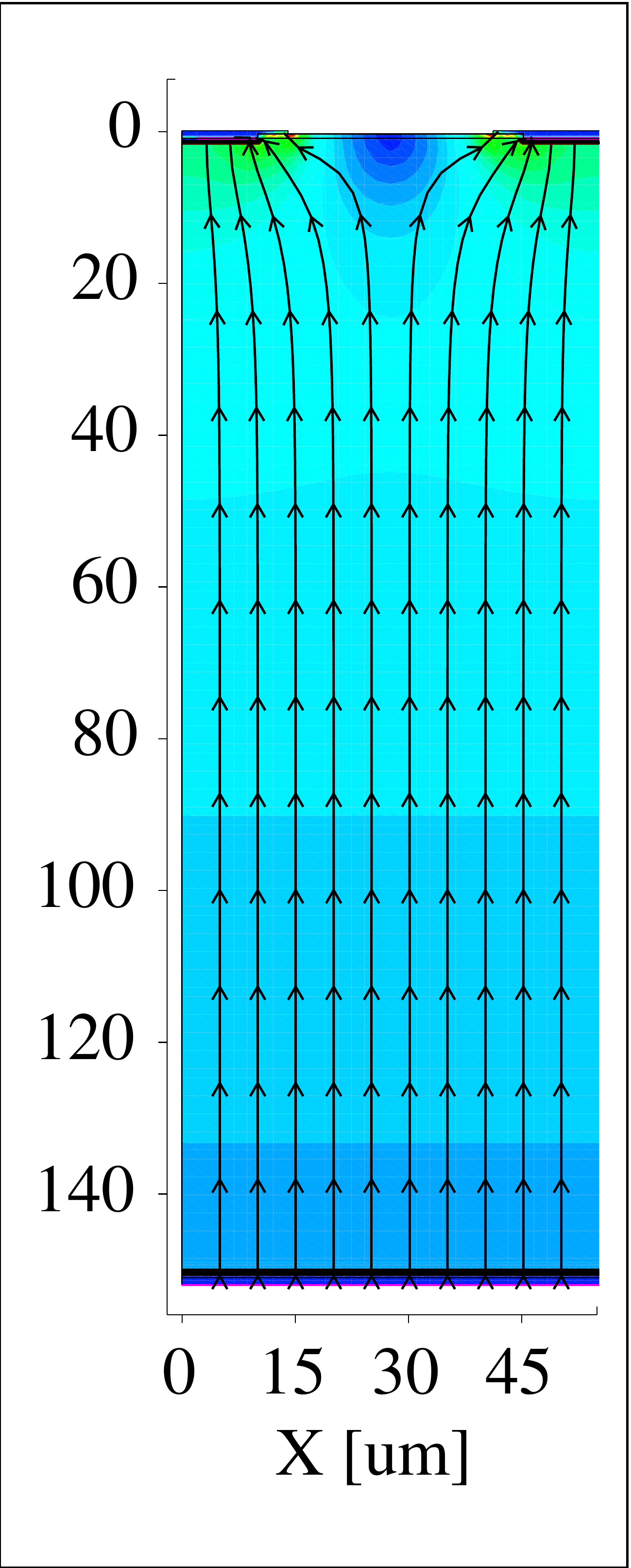}\put(-80,-10){b)}\put(-45,-10){\small X [$\muup$m]}
  \includegraphics[width = 0.03\textwidth ]{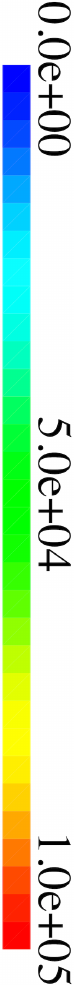}\put(0, 135){\rotatebox{-90}{\scriptsize Electric Field [$\mathrm{V \cdot cm^{-1}}$]}}\hfill
  \includegraphics[trim=0.5cm 3.0cm 0.5cm 2.1cm, width = 0.18\textwidth, clip]{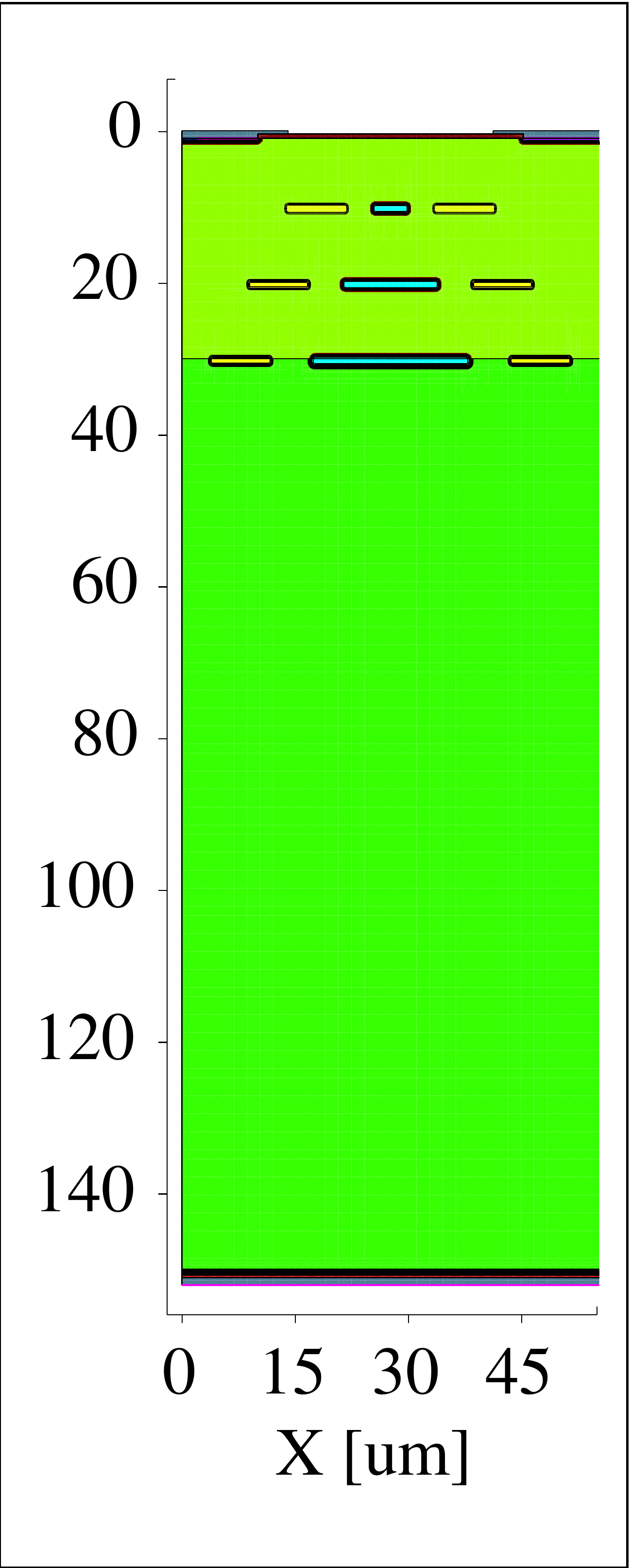}\put(-80,-10){c)}\put(-56, 175){$\mathrm{RO_{l}}$}\put(-17, 175){$\mathrm{RO_{r}}$}\put(-45,-10){\small X [$\muup$m]}
    \includegraphics[width = 0.025\textwidth ]{neladleg.pdf}\put(0, 145){\rotatebox{-90}{\scriptsize Doping Concentration [$\mathrm{cm^{-3}}$]}}\hfill
  \includegraphics[trim=0.5cm 3.0cm 0.5cm 2.1cm, width = 0.18\textwidth, clip]{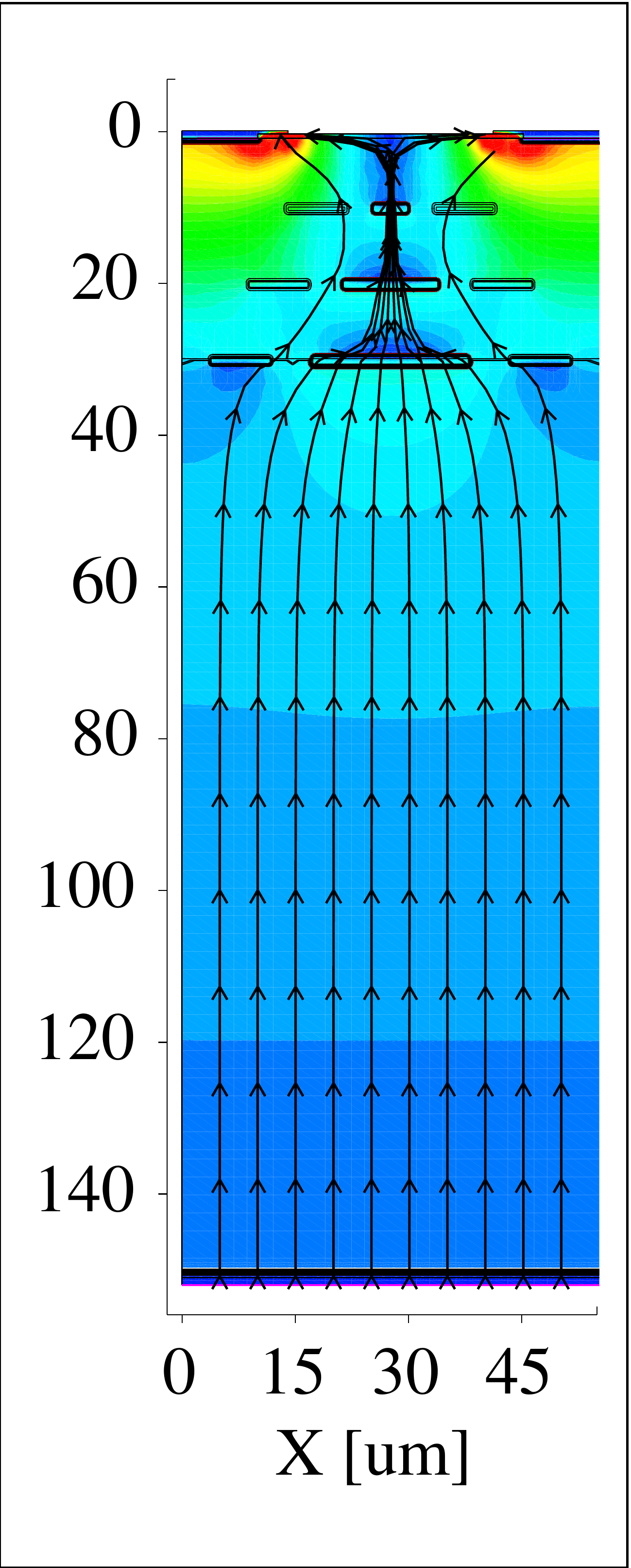}\put(-80,-10){d)}\put(-45,-10){\small X [$\muup$m]}
  \includegraphics[width = 0.03\textwidth ]{nelad_ef_leg.pdf}\put(0, 135){\rotatebox{-90}{\scriptsize Electric Field [$\mathrm{V \cdot cm^{-1}}$]}}\hfill
  \caption{
The doping concentration profiles of a) a standard planer sensor and c) the n-ELAD sensor with $n\mathrm{_{di}} = 2.8\mathrm{\cdot10^{15}\,cm^{-3}}$ at $U=280$\,V are shown. 
RO$_{\textrm{l}}$ (RO$_{\textrm{r}}$) denote the left (right) readout electrode (omitted hereafter).
The resulting electric field strength profiles are depicted in b) and d).
}
  \label{fig:el}
\end{figure}

Two types of ELAD sensors have been designed -- p-in-n (n-ELAD) and n-in-p (p-ELAD~\cite{elad}) sensors.
The n-ELAD sensor is formed by an n-type substrate in combination with three n-type epitaxial layers (resulting from the production process),
 three interjacent buried implant layers, each containing one p-implant with its centre in the middle between two readout implants and two adjacent n-implants,
 p-type readout implants and a backplane implant.
The design of the n-ELAD sensor is shown in Fig.~\ref{fig:el}~c), where RO$_{\textrm{l}}$ and RO$_{\textrm{r}}$ denote the left and the right readout electrode.
For comparison, a standard planar n-type sensor is shown in a).
Yellow (blue) areas represent buried n-implants (p-implants), with a light-green (dark-green) colour showing the epitaxial zone (substrate) of the sensor. 
The distance from the top of the sensor to the first buried implant layer is 10\,$\muup$m, the distance between buried implant layers is 10\,$\muup$m, the total thickness of the epitaxial layer amounts to 30\,$\muup$m.
Analogously, the p-ELAD sensor is formed by a p-type substrate and three p-type epitaxial layers,
 with three buried implant layers with an inverted doping in each layer with respect to the n-ELAD,
 n-type readout implants, a moderated p-spray isolation and a backplane implant.
The pitch for both types of the ELAD is 55\,$\muup$m in order to match the TimePix3 readout chip footprint~\cite{tp3}; the readout implant size amounts initially to 20\,$\muup$m.

The Technology Computer-Aided Design (TCAD) framework by SYNOPSYS is used for the simulation presented herein~\cite{syn}.
Two types of simulations have been performed: electric field simulations and transient simulations with a minimum ionising particle (MIP) traversing the sensor.
Simulations of the electric field have been performed to describe the behaviour of the ELAD sensors in 2D and 3D.
In order to study the performance of ELAD sensors, transient simulations of 2D geometries have been executed for n-type and p-type ELAD sensors.
This allows for the analysis of the distribution of the collected charge on neighbouring readout electrodes as a function of the impact position, i.e.\ the $\etaup$-function.
The parameters which have been optimised to approach a linear behaviour of the $\etaup$-function are: the depth of the first buried implant layer w.r.t.\ the sensor surface, the distance between buried implant layers,
 the readout implant size, the bias voltage and the doping concentration of the buried implants. 
The electric field profile yielding an optimal charge sharing behaviour is obtained via a scan over the specified parameters.
Here, we detail the optimisation process with respect to the buried implant concentration and the readout implant size.

\subsection{Electric field}
\label{sec:ef}
The electric field simulations are performed in a quasi-stationary mode, in which the voltage is applied to the electrodes of a device.
The voltage is ramped incrementally to a target value; in each step the electric field profile is calculated
 by solving the Poisson's equation with the charge density from the previous iteration as starting value.

In Fig.~\ref{fig:el} the total electric field with its electric field lines is presented for b) a standard, planar n-type sensor and d) the n-ELAD sensor
 with a total thickness of 150~$\muup$m and a thickness of the epitaxial layer of 30~$\muup$m.
The electric field lines in the standard planar n-type sensor lead directly to the readout electrodes. 
In the n-ELAD sensor, the electric field lines change their behaviour with respect to the standard planar sensor, effectively pointing to the centre between two readout electrodes. 
Hence, the buried implants create a lateral electric field component in the bulk of the sensor rendering a position-dependent charge sharing possible.
The areas of larger field strength in vicinity of the readout electrodes in case of the n-ELAD compared to the standard, planar sensor is a result of the higher background concentration in the epitaxial part. 

\begin{figure}[t!]
 \centering
  \includegraphics[trim=0.5cm 3.0cm 0.5cm 2.1cm,, width = 0.18\textwidth, clip]{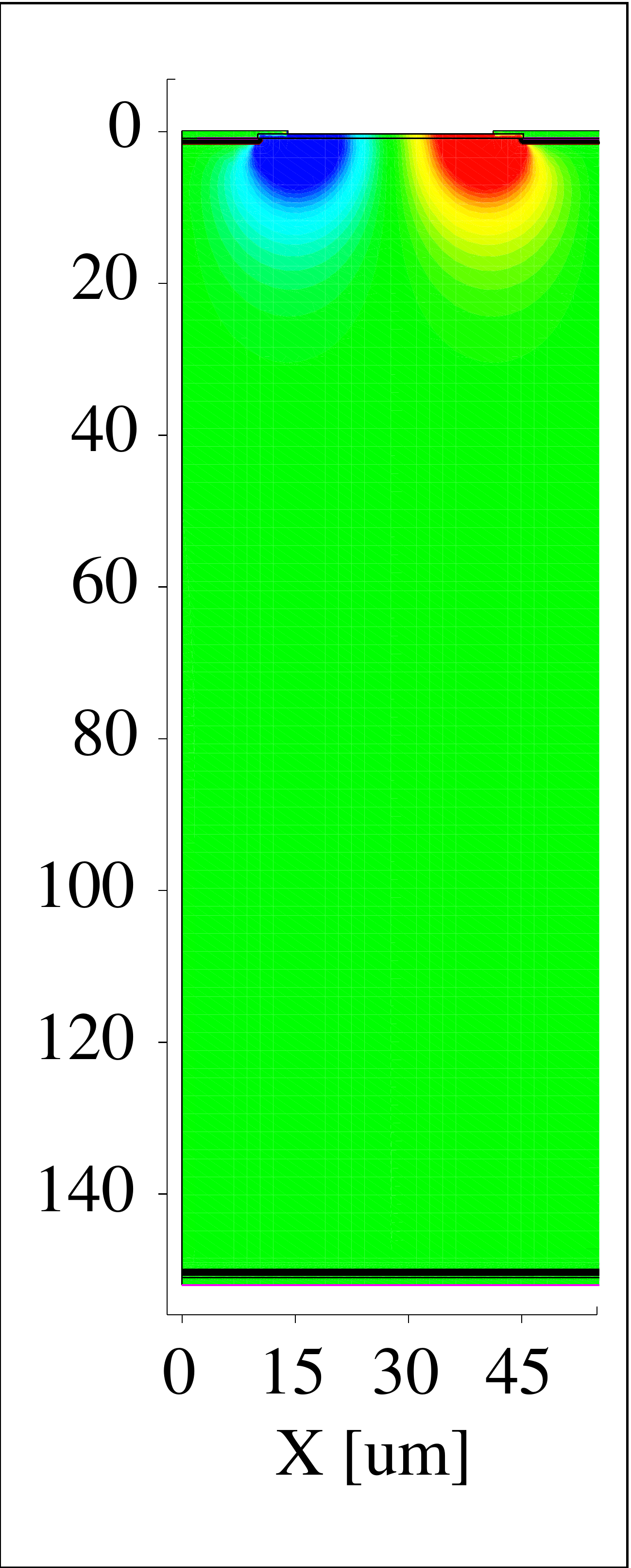}\put(-70,-10){a)}\put(-88,100){\rotatebox{90}{\small Y [$\muup$m]}}\put(-44,-10){\small X [$\muup$m]}
  \includegraphics[trim=0.5cm 3.0cm 0.5cm 2.1cm,, width = 0.18\textwidth, clip]{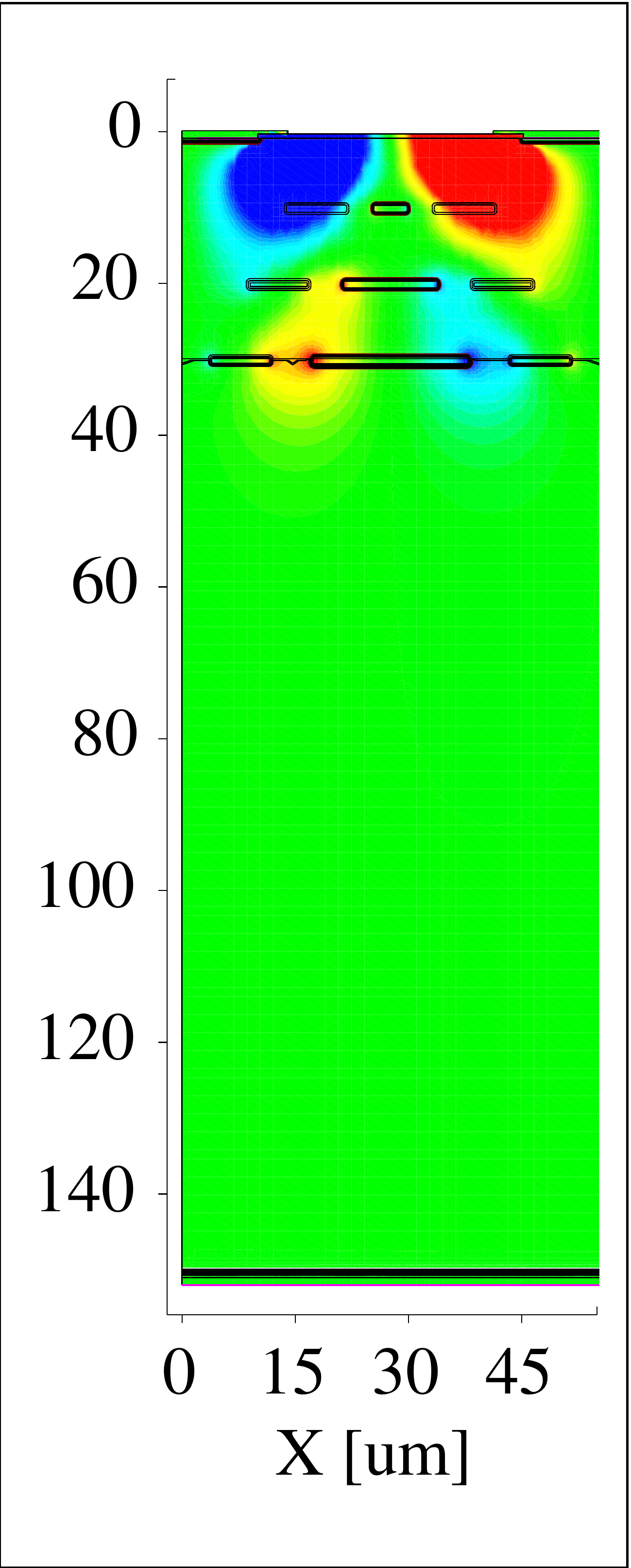}\put(-70,-10){b)}\put(-44,-10){\small X [$\muup$m]}
  \includegraphics[trim=0.5cm 3.0cm 0.5cm 2.1cm,, width = 0.18\textwidth, clip]{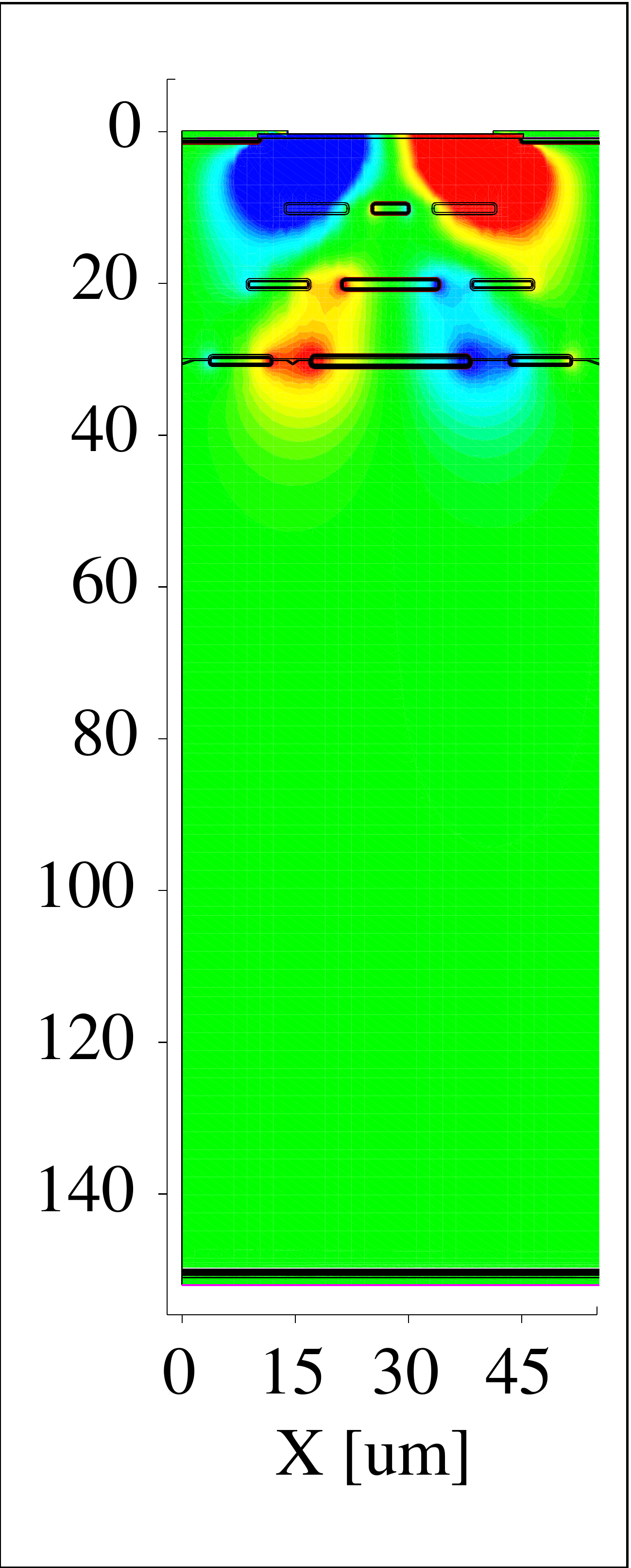}\put(-70,-10){c)}\put(-44,-10){\small X [$\muup$m]}
  \includegraphics[trim=0.5cm 3.0cm 0.5cm 2.1cm,, width = 0.18\textwidth, clip]{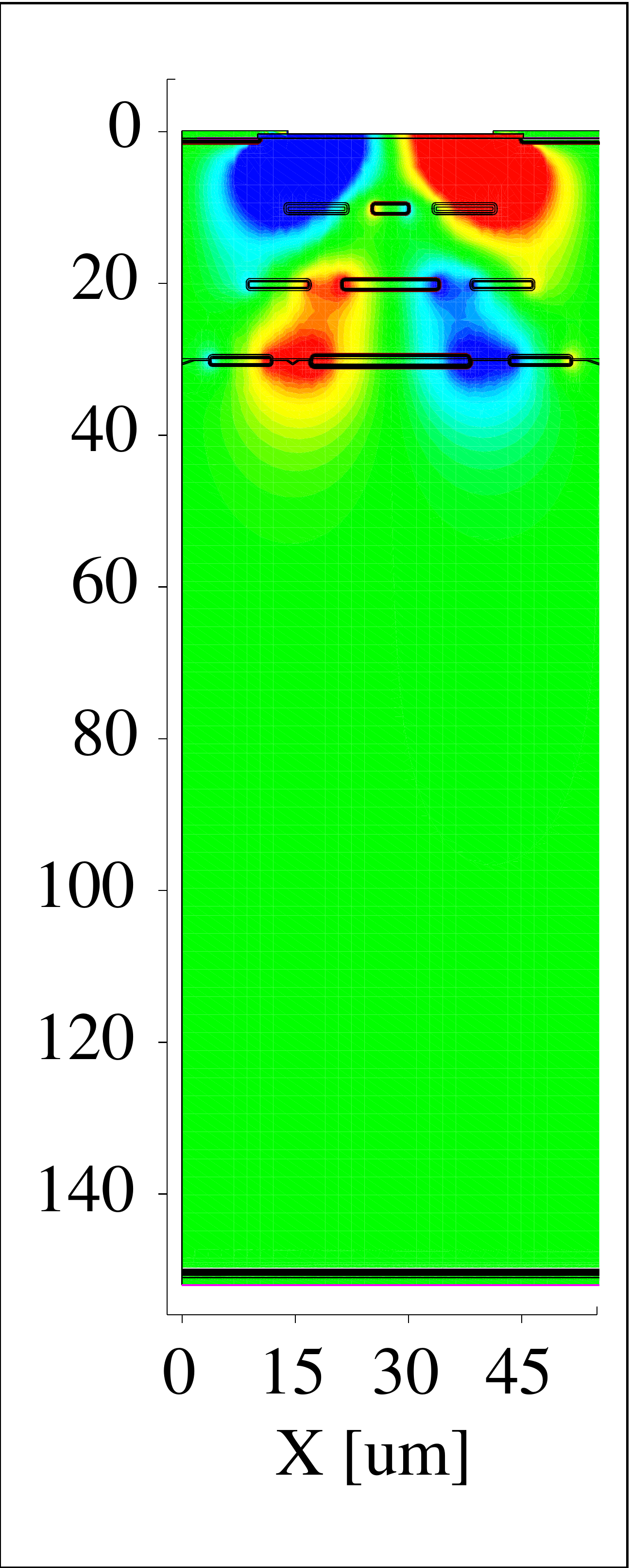}\put(-70,-10){d)}\put(-44,-10){\small X [$\muup$m]}
  \includegraphics[trim=0.5cm 3.0cm 0.5cm 2.1cm,, width = 0.18\textwidth, clip]{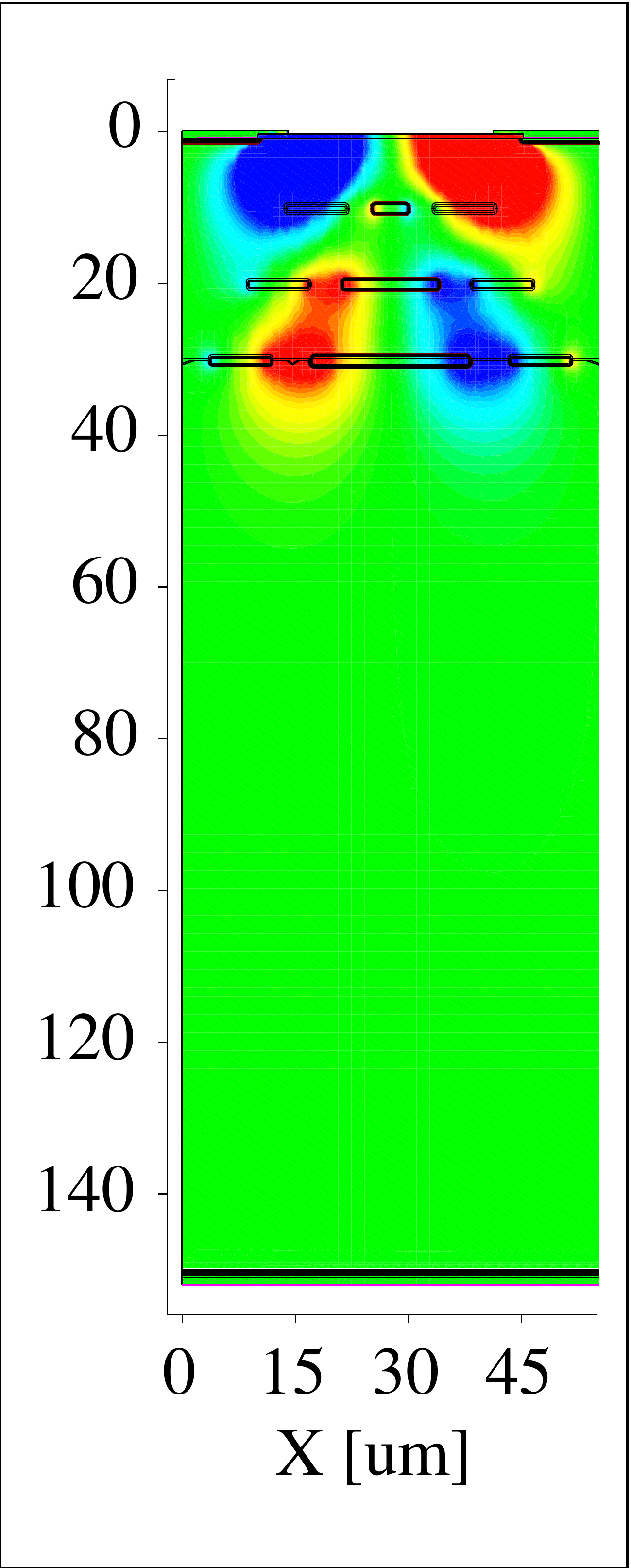}\put(-70,-10){e)}\put(-44,-10){\small X [$\muup$m]}
  \includegraphics[height=6.1cm ]{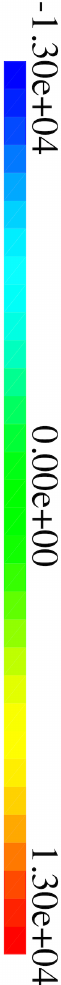}\put(5, 135){\rotatebox{-90}{\footnotesize Electric Field [$\mathrm{V \cdot cm^{-1}}$]}}
  \caption{
The lateral electric field component for a) a standard planar n-sensor and b) - e) the n-ELAD sensors with increasing buried implant concentrations (see text) at $U=280$\,V is shown.
}
  \label{fig:ef}
\end{figure}
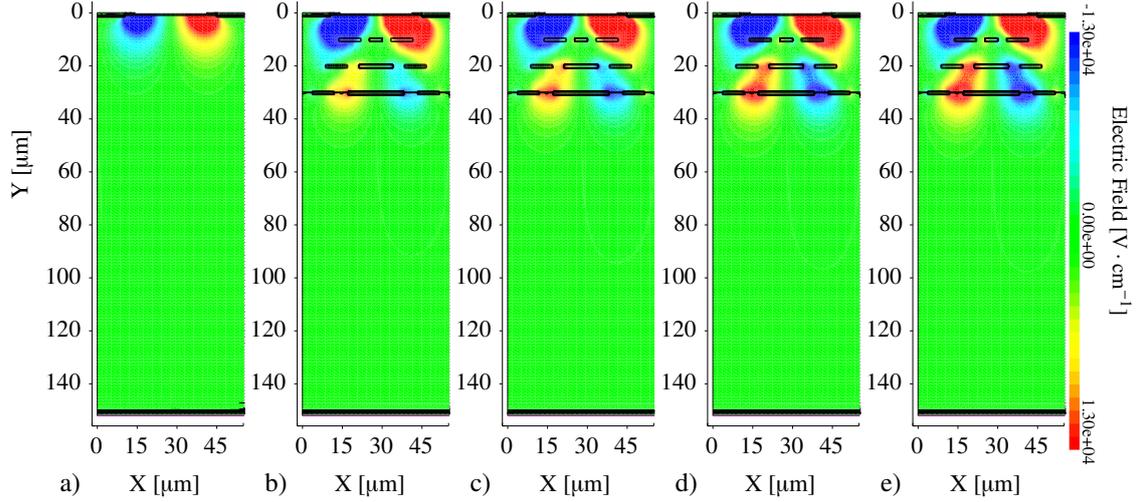

Figure~\ref{fig:ef} shows the electric field profile in the lateral direction for a) a standard, planar n-type sensor
 and b) - e) the n-ELAD sensor with different buried implant concentrations
  $n\mathrm{_{di}}$ of 2.0$\mathrm{\cdot10^{15}\,cm^{-3}}$, 2.4$\mathrm{\cdot10^{15}\,cm^{-3}}$, 2.8$\mathrm{\cdot10^{15}\,cm^{-3}}$ and 3.0$\mathrm{\cdot10^{15}\,cm^{-3}}$, respectively.
Red (blue) areas indicate a hole drift component to the right (left).
Equivalently in the case of the p-ELAD sensors, the electron drift paths are influenced by the buried implants.

The buried implants create a defined electric field in the lateral direction. 
With increasing doping concentration of the buried implants the lateral electric field becomes stronger and its effect on the drift paths of the charge carriers larger.
The electric field in the longitudinal direction is created by the applied voltage. 
The ratio between the longitudinal and lateral components of the electric field changes with the applied voltage and buried implant concentrations.
Identification of the optimal operational voltage allows for an effective usage of the ELAD sensor, as is detailed in Sec.~\ref{sec:tr}.

\begin{figure}[t!]
  \centering
  \includegraphics[trim=0.2cm 0.5cm 0.5cm 0.5cm, height=6.1cm, clip]{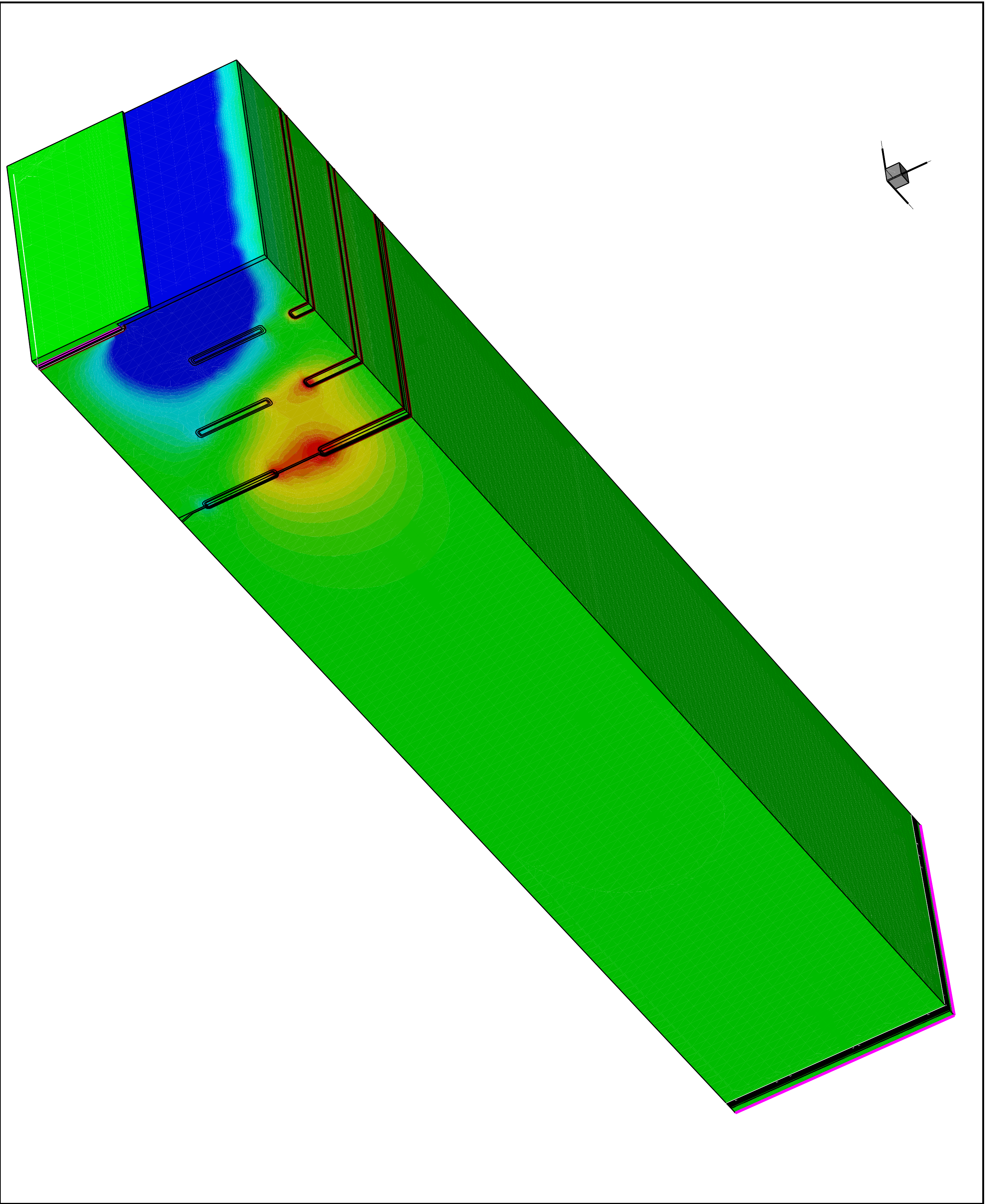}\put(-110,-10){a)}
  \includegraphics[height=6.5cm ]{lat_ef_leg.pdf}\put(5, 150){\rotatebox{-90}{\footnotesize Electric Field [$\mathrm{V \cdot cm^{-1}}$]}}
  \hfill 
  \includegraphics[trim=3cm 3.5cm 0.5cm 0.5cm, height=5cm, clip]{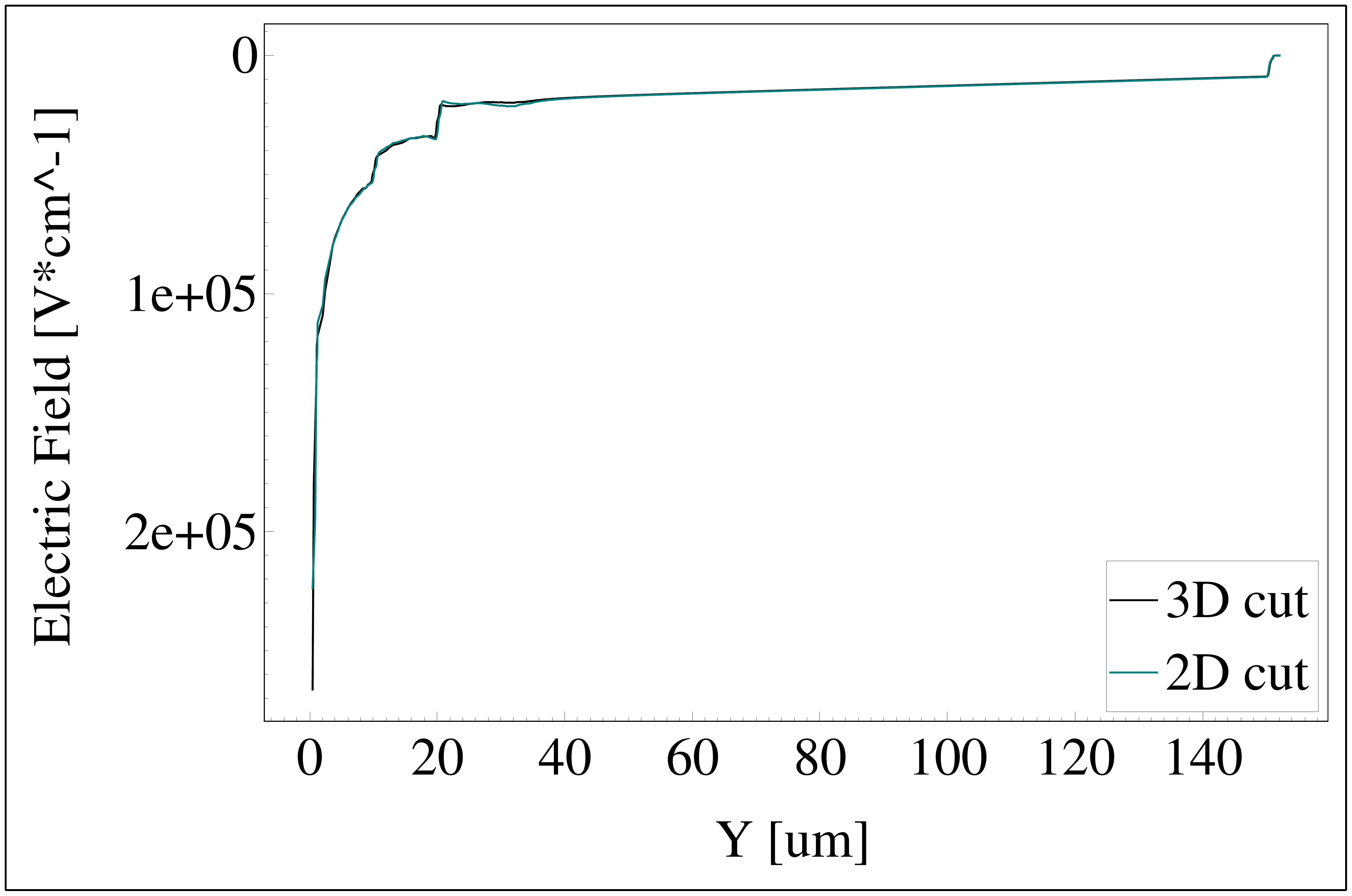}\put(-250,-10){b)}\put(-235, 35){\rotatebox{90}{ \small Electric Field [$\mathrm{V \cdot cm^{-1}}$]}}\put(-110,-10){\small Y [$\muup$m]}
  \hfill 
  \caption{
The electric field in the lateral direction in the n-ELAD sensor resulting from a 3D simulation is shown for a buried implant concentration of $n\mathrm{_{di}} = 2.8\mathrm{\cdot10^{15}\,cm^{-3}}$ with a strip readout. 
In b) the field strength along the depth of the sensor is superimposed for the 2D and 3D simulation at $x_0 = 10\,\muup$m, $z_0 = 20\,\muup$m.}
  \label{fig:3d}
\end{figure}

3D electric fields simulations with strip and pixel readouts have been carried out to verify the 2D electric field simulations.
The result of the 3D simulation for the strip readout is shown in~Fig.~\ref{fig:3d}~a).
In order to optimise the duration of the simulation process a quarter of the unit cell has been modelled exploiting the symmetry of the design.
The cross section of the 3D electric field profile resembles the one obtained from the 2D simulations, compare Fig.~\ref{fig:ef}~(d). 
In Fig.~\ref{fig:3d}~b) the electric field strength for the 2D and the 3D simulation are superimposed for a cut line along the depth of the sensor at a given position ($x_0,z_0$). 
No significant deviation is observed.

\subsection{Transients}
\label{sec:tr}
In transient simulations the response of the device to a traversing particle is simulated. 
The transient simulation approximates a continuous process as a sequence of frames with a short time difference. 
In each frame and for each mesh node the Poisson's equation and carrier continuity equations are solved for electrons and holes.
In our simulation, a four-unit-cells geometry has been used with the region of interest extending from the centre of the second readout implant towards the next rightwards unit cell boundary.  
MIP incident positions have been simulated at 0\,$\muup$m (centre of the \textit{left} readout implant), 6.3\,$\muup$m, 11.6\,$\muup$m, 16.9\,$\muup$m, 22.2\,$\muup$m and 27.5\,$\muup$m (unit cell boundary).

\begin{figure}[t!]
  \centering
  \includegraphics[trim=0.5cm 2.2cm 0.5cm 2.1cm, clip, width=0.18\textwidth]{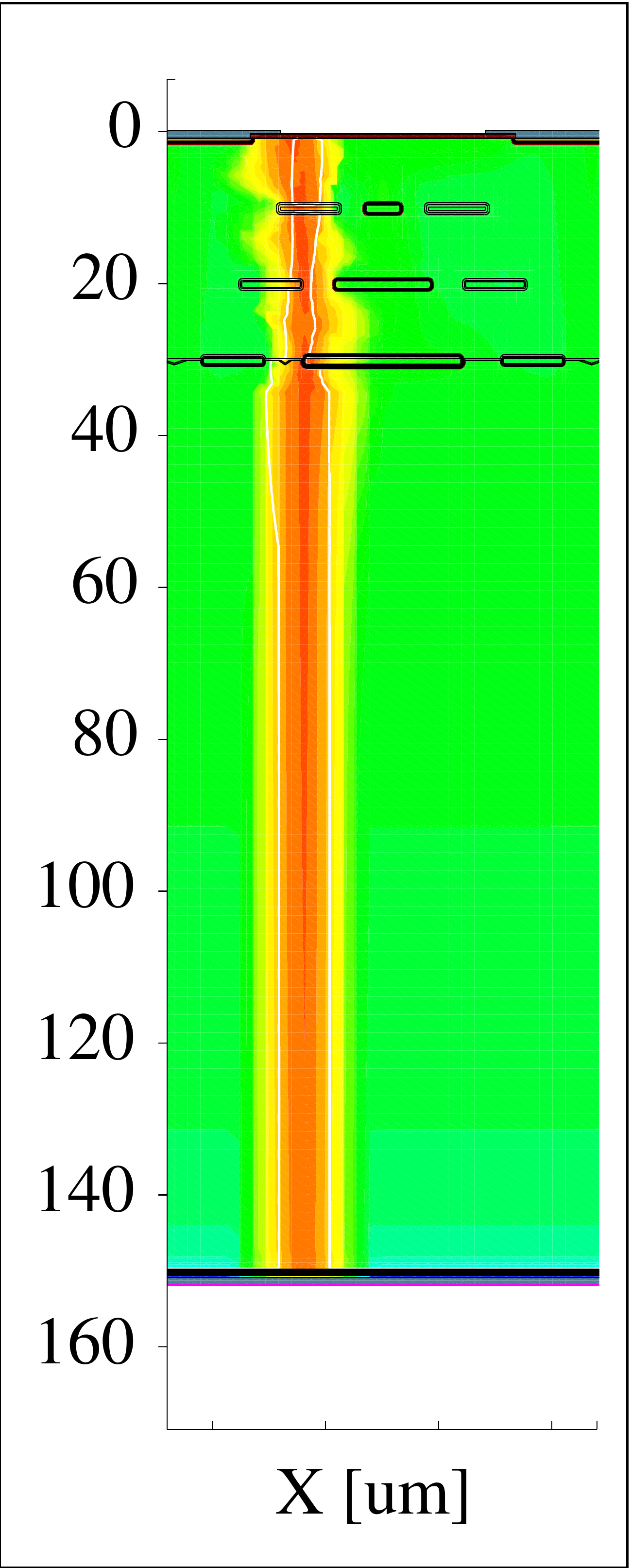}\put(-60, -30){a) $t=0.0$\,ns}\put(-53, 10){\small5}\put(-42, 10){\small20}\put(-26, 10){\small35}\put(-12, 10){\small50}\put(-88,100){\rotatebox{90}{\small Y [$\muup$m]}}\put(-44,-10){\small X [$\muup$m]}\hfill
  \includegraphics[trim=0.5cm 2.2cm 0.5cm 2.1cm, clip, width=0.18\textwidth]{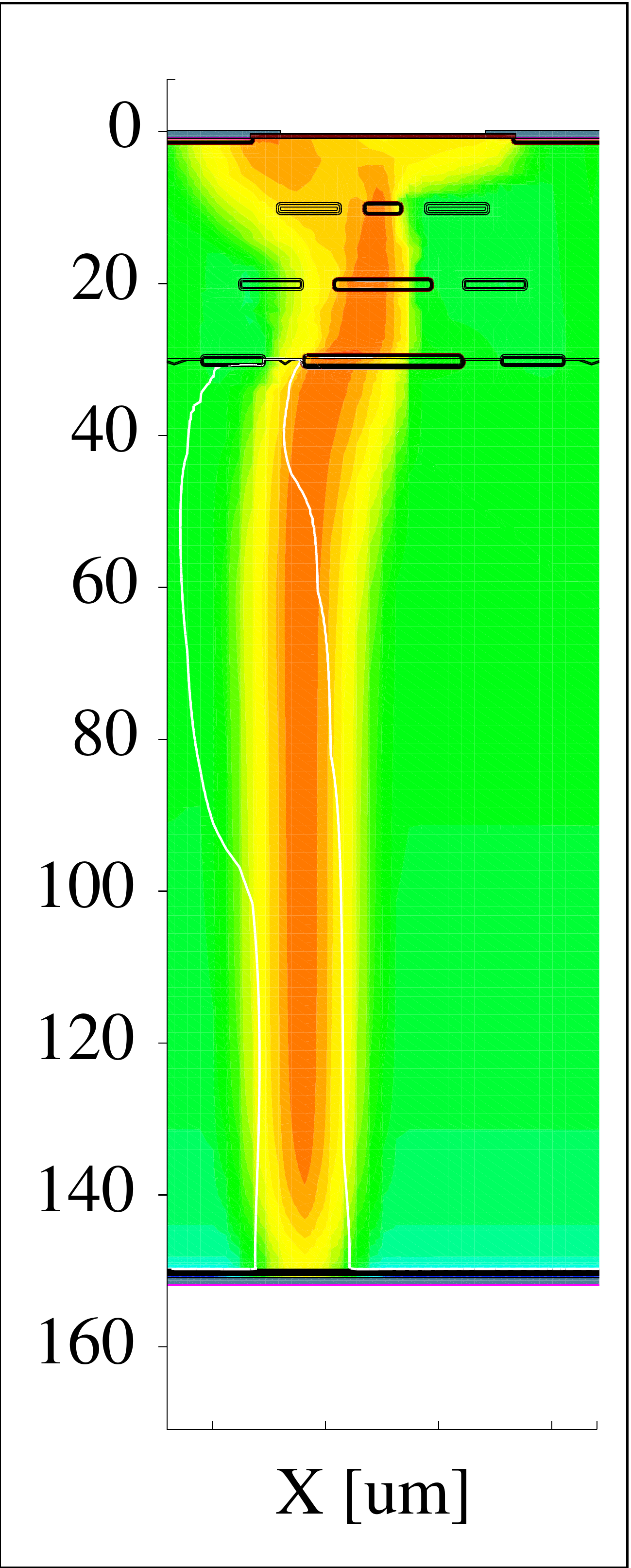}\put(-60, -30){b) $t=0.4$\,ns}\put(-53, 10){\small5}\put(-42, 10){\small20}\put(-26, 10){\small35}\put(-12, 10){\small50}\put(-44,-10){\small X [$\muup$m]}\hfill
  \includegraphics[trim=0.5cm 2.2cm 0.5cm 2.1cm, clip, width=0.18\textwidth]{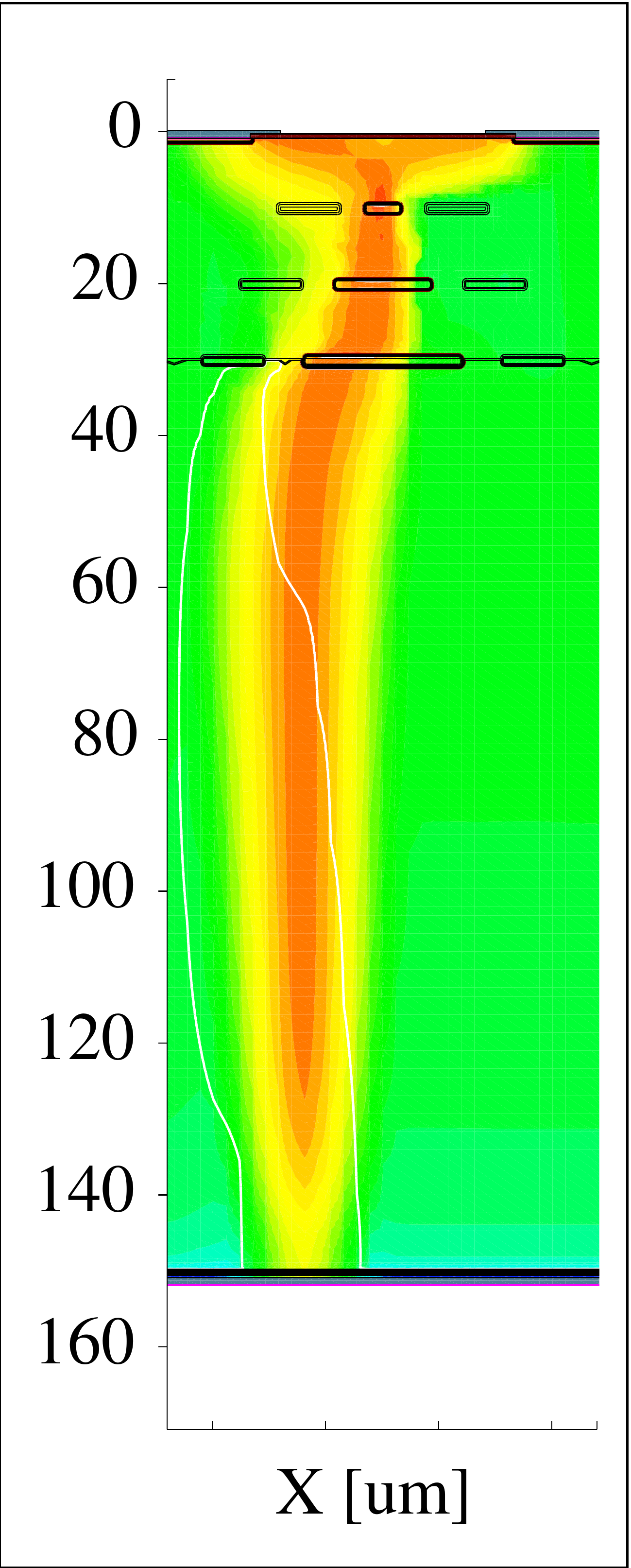}\put(-60, -30){c) $t=0.8$\,ns}\put(-53, 10){\small5}\put(-42, 10){\small20}\put(-26, 10){\small35}\put(-12, 10){\small50}\put(-44,-10){\small X [$\muup$m]}\hfill
  \includegraphics[trim=0.5cm 2.2cm 0.5cm 2.1cm, clip, width=0.18\textwidth]{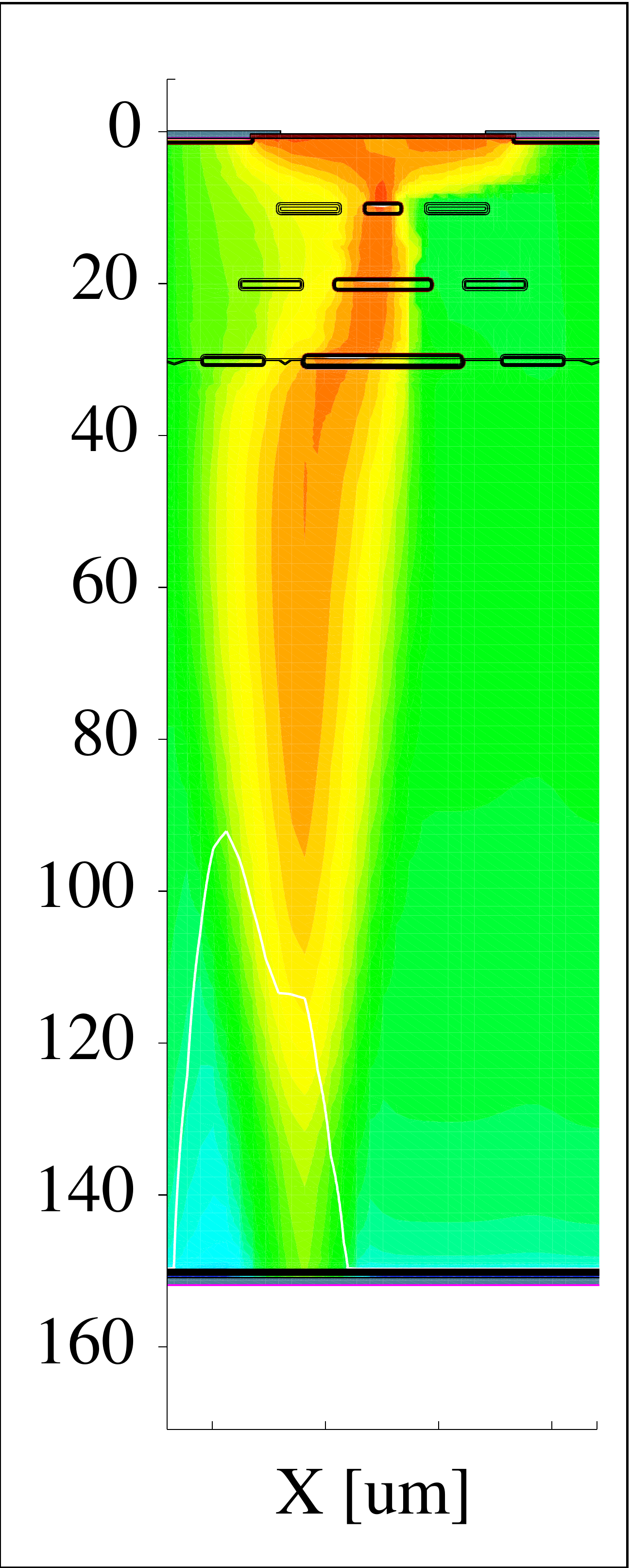}\put(-60, -30){d) $t=2.0$\,ns}\put(-53, 10){\small5}\put(-42, 10){\small20}\put(-26, 10){\small35}\put(-12, 10){\small50}\put(-44,-10){\small X [$\muup$m]}\hfill
  \includegraphics[trim=0.5cm 2.2cm 0.5cm 2.1cm, clip, width=0.18\textwidth]{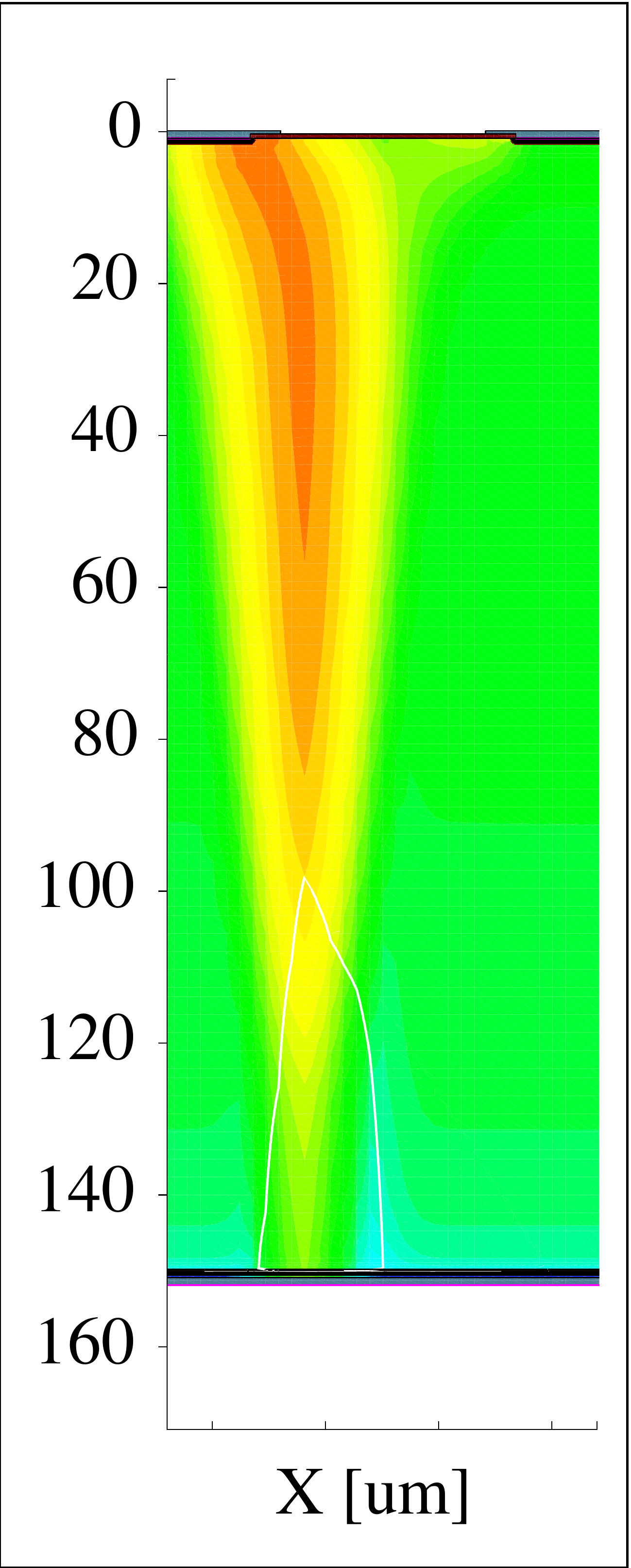}\put(-60, -30){e) $t=2.0$\,ns}\put(-53, 10){\small5}\put(-42, 10){\small20}\put(-26, 10){\small35}\put(-12, 10){\small50}\put(-44,-10){\small X [$\muup$m]}
  \includegraphics[ height=6.5cm]{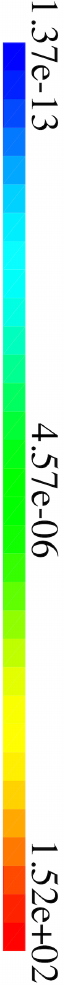}\put(0, 160){\rotatebox{-90}{\footnotesize Hole Current Density [$\mathrm{A \cdot cm^{-2}}$]}}
  \caption{
The hole current density for the n-ELAD sensor with $n\mathrm{_{di}} = 2.8\mathrm{\cdot10^{15}\,cm^{-3}}$ is shown in a) - d) at four different points in time at $U=280$\,V for a MIP incident at 16.9\,$\muup$m.
In e), the hole current density is shown at $t=2\,$ns for a standard planer n-type sensor.
}
 \label{fig:tr}
\end{figure}
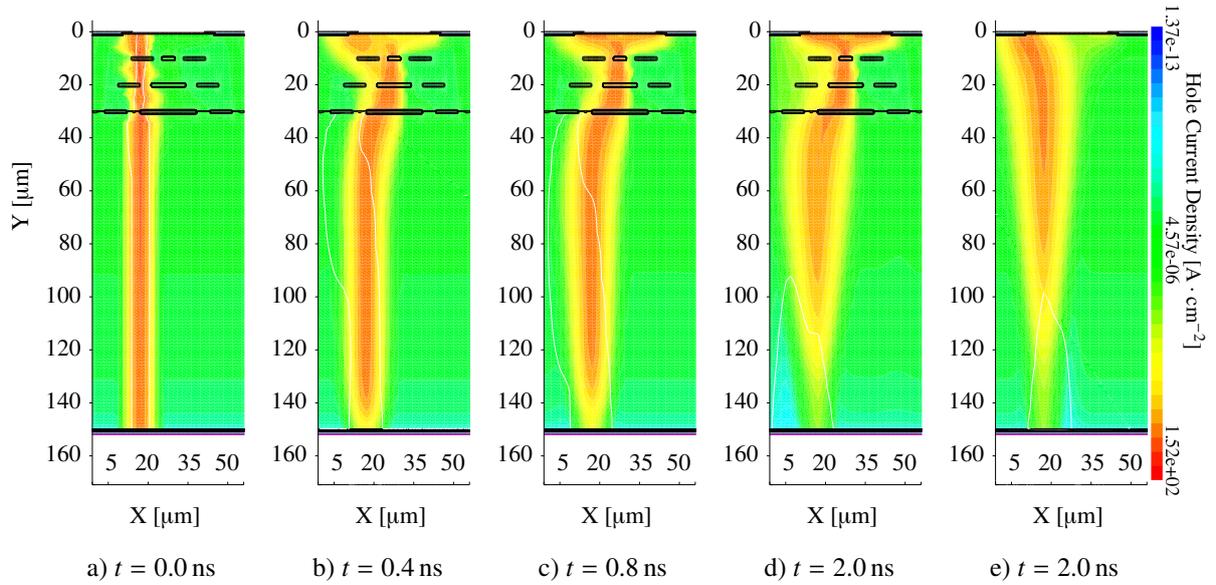

In Fig.~\ref{fig:tr} the hole current density in the n-ELAD sensor with $n\mathrm{_{di}} = 2.8\mathrm{\cdot10^{15}\,cm^{-3}}$ is shown for four time steps at a MIP incident position of 16.9\,$\muup$m.
In the first two time steps (Fig.~\ref{fig:tr}~a) and b)) the charge carriers created above the buried implants are collected by the nearest, i.e.\ the left electrode.
Charges created beneath the buried implants change their drift path and move towards the centre (Fig.~\ref{fig:tr}~c) and d)) reducing the required diffusion length to cross the cell boundary.
A considerable fraction of the created charge is eventually collected at the neighbouring electrode. 
Congruent results have been obtained for the p-ELAD sensor.
For comparison, Fig.~\ref{fig:tr}~e) shows a standard planar sensor at $t = 2.0\,$ns with the MIP traversing at the same position with only negligible charge sharing.
In the following part, we exemplify the optimisation process for two of the ELAD parameters at the example of the buried layer concentration and the readout implant size. 

\begin{figure}[t!]
  \centering
  \includegraphics[trim=0.1cm 0cm 1.cm 1.5cm, width = 0.48\textwidth, clip]{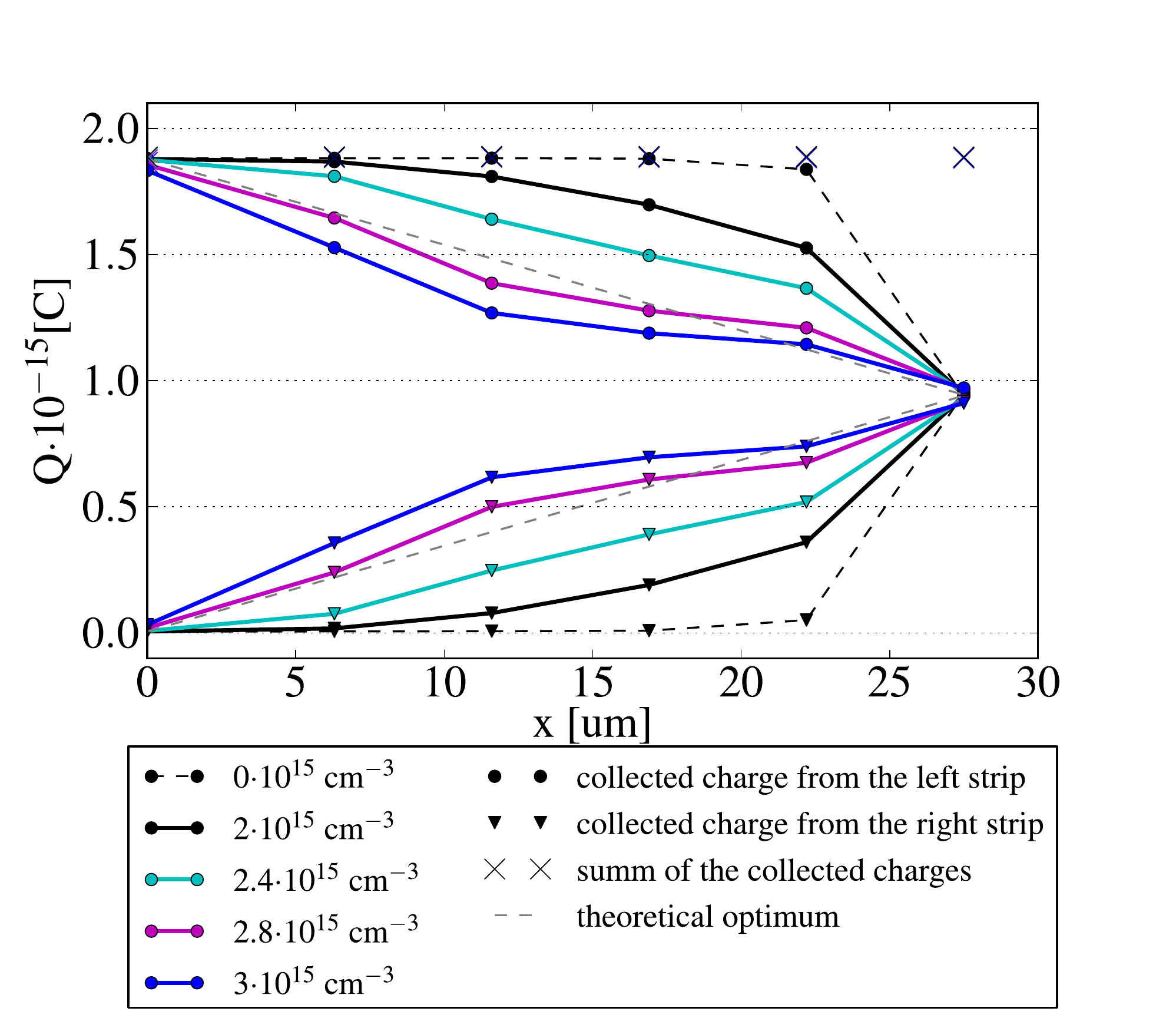}\put(-220,170){a)} \hfill
  \includegraphics[trim=0.1cm 0cm 1.cm 1.5cm, width = 0.48\textwidth, clip]{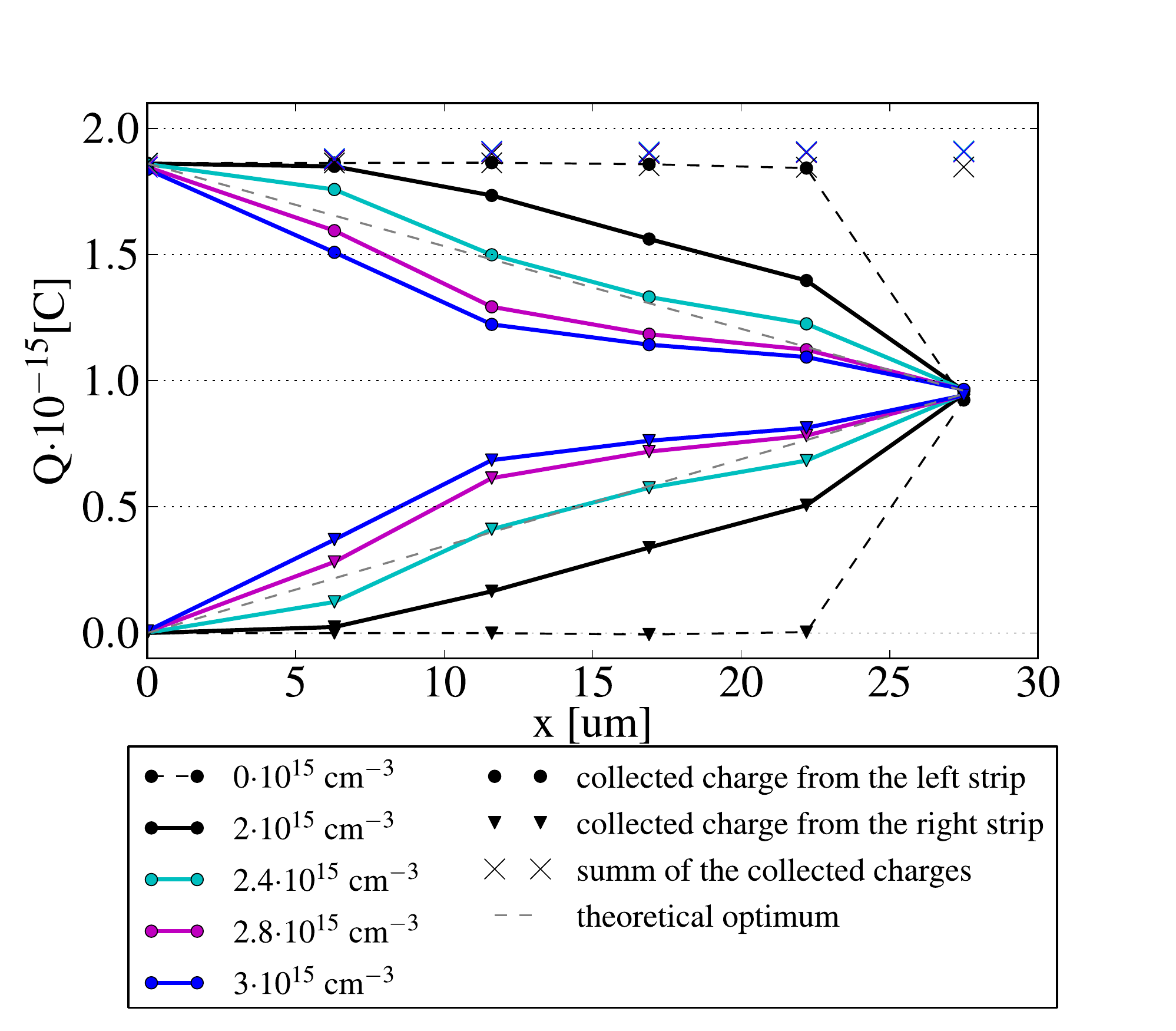}\put(-220,170){b)}
  \caption[]{
The $\etaup$-function for different buried implant concentrations for a) the n-ELAD sensor at $U=280$\,V and b) the p-ELAD sensor at $U=300$\,V. 
Lines connecting data points guide the eye.
}
  \label{fig:eta}
\end{figure}

In Fig.~\ref{fig:eta} the collected charge from two neighbouring readout electrodes as a function of the six different MIP positions ($\etaup$-function) is presented.
The response of the n-ELAD and the p-ELAD sensor with different buried implant concentrations are indicated as
 circles (triangles) showing the collected charge from the left (right) strip
 and as crosses corresponding to the sum of the collected charges. 
For comparison, a sensor with an epitaxial layer but without buried implants (\textit{empty} ELAD) is added. 
The trend of the $\etaup$-functions depends on the buried implant concentration.
In the case of the n-ELAD (Fig.~\ref{fig:eta}~a)) the lowest simulated buried implant concentration of 2.0$\mathrm{\cdot10^{15}\,cm^{-3}}$ (black~line) does not yield an optimal $\etaup$-function.
However, the effect of the buried implants is already noticeable w.r.t.\ the empty ELAD (black dots and black dashed lines). 
The blue line shows the result for the highest buried implant concentration of 3.0$\mathrm{\cdot10^{15}\,cm^{-3}}$. 
In this case, the buried implants create a too strong electric field in the lateral direction yielding an excessive charge sharing.
For an applied voltage of 280\,V the optimal buried implant concentration is 2.8$\mathrm{\cdot10^{15}\,cm^{-3}}$ approaching the linear behaviour closest for the concentrations under study.
For the p-ELAD sensor and an applied voltage of 300\,V (see Fig.~\ref{fig:eta}~b)) the optimal buried implant concentration amounts to 2.4$\mathrm{\cdot10^{15}\,cm^{-3}}$.
Hence, for a given concentration the operational voltage should be carefully chosen to yield an optimal ratio between the electric field strength in the longitudinal and the lateral direction. 
If the applied voltage is too low (too high) the effect of the charge sharing by the buried implants is too strong (too weak) resulting in a considerable deviation of the $\etaup$-function from the linear case.

\begin{figure}[t!]
  \centering
  \includegraphics[trim={0.1cm 0cm 1.cm 1.5cm}, width = 0.48\textwidth, clip]{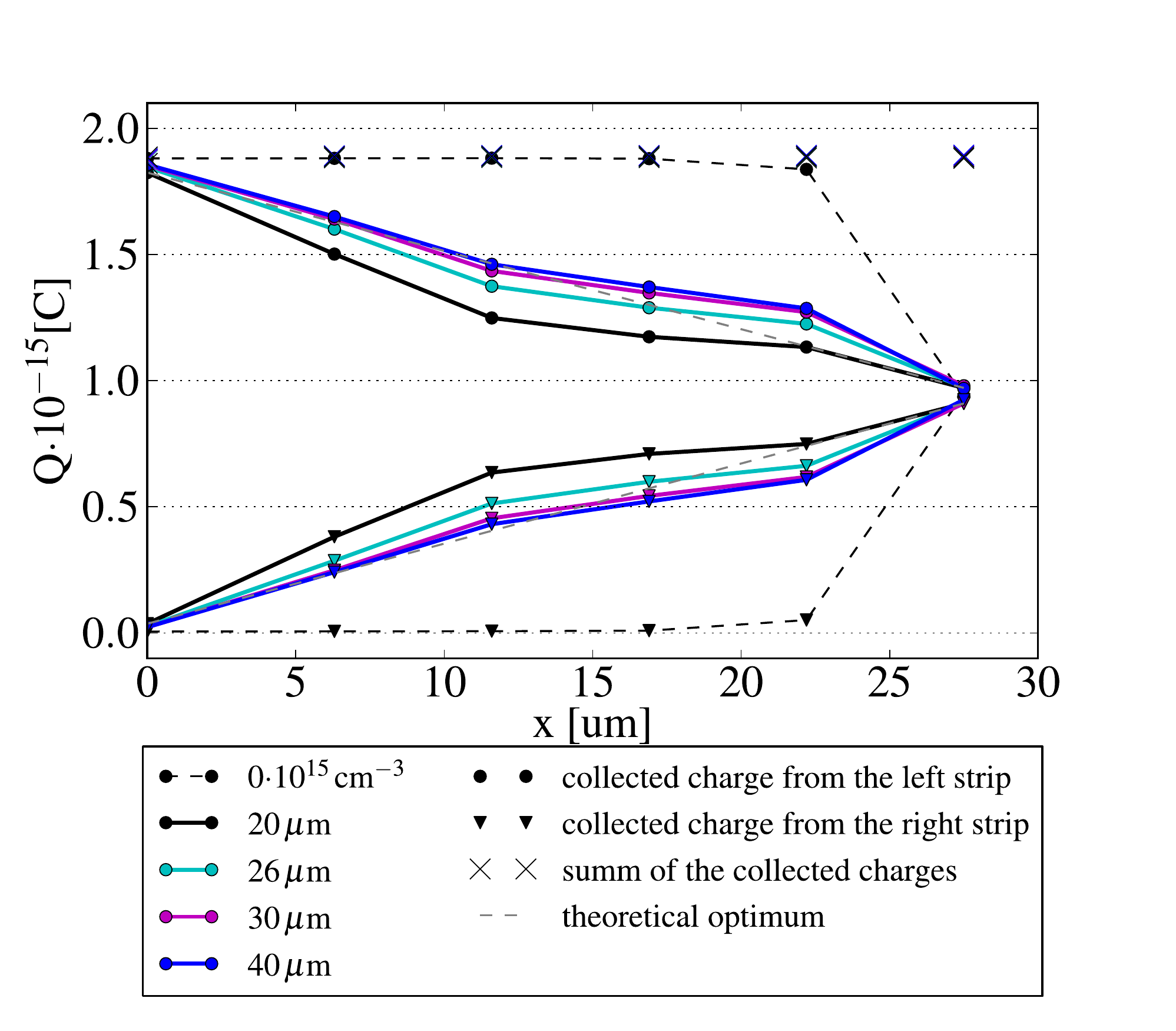}\put(-220,170){a)} \hfill
  \includegraphics[trim={0.1cm 0cm 1.cm 1.5cm}, width = 0.48\textwidth, clip]{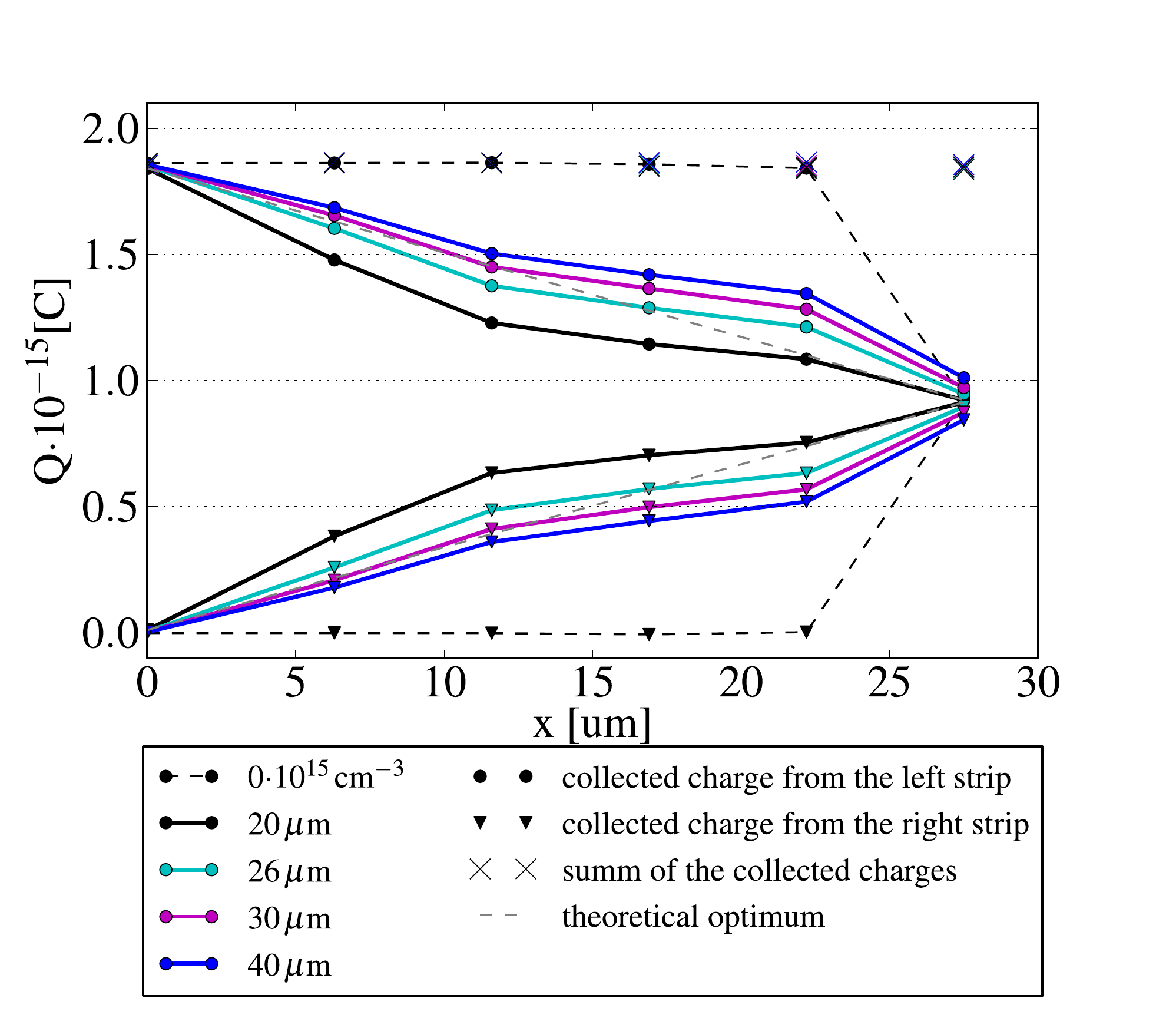}\put(-220,170){b)}
  \caption[]{
The $\etaup$-function for different sizes of the readout implant for a) the n-ELAD sensor at $U=275$\,V and b) the p-ELAD sensor at $U=300$\,V.
Lines connecting data points guide the eye.
}
  \label{fig:rosize}
\end{figure}

To study the effect of the readout implant size on the $\etaup$-function several readout implant sizes have been simulated at $n\mathrm{_{di}} = 3.0\mathrm{\cdot10^{15}\,cm^{-3}}$. 
The bias voltage in the n-ELAD and p-ELAD sensors is 275\,V and 300\,V, respectively. 
In Fig.~\ref{fig:rosize} the transient simulation results for a) the n-ELAD and b) the p-ELAD sensor are shown.
The simulated readout implant sizes are 20\,$\muup$m (black continuous lines), 26\,$\muup$m, 30\,$\muup$m and 40\,$\muup$m (blue lines).
The black dashed lines show the empty n-ELAD and p-ELAD sensors with zero buried implant concentration and readout implant size of 20\,$\muup$m.
For both the n-ELAD and the p-ELAD sensors the size of the readout implant impacts the shape of the $\etaup$-function.
With increasing size of the readout implant the effect of the charge sharing decreases.  
Thus, the size of the readout cell should be carefully chosen w.r.t.\ $n_{\mathrm{di}}$ and the applied bias voltage to tune the shape of the $\etaup$-function in ELAD sensors.

\section{ELAD production process}
\label{sec:pr}
The realisation of the ELAD sensors requires a new production process combining several repetitive steps.
Each step includes ion beam implantations and epitaxial silicon growth similar to buried implants in buried collector technology. 
At first, the bottommost layer of buried implants is implanted on a silicon wafer followed by an epitaxial layer that is grown on top of the implanted silicon. 
The process is repeated three times. 
Finally, after the last epitaxial growth, common processes used for standard planar sensors are applied e.g.\ edging and preparation of readout implants.

As the epitaxial growth process imposes a considerable temperature budget on the implants, process simulations have been executed in order to quantify a possible effect on the charge sharing mechanism. 
The process simulation of a buried Boron implant shows a change in size of about 1\,$\muup$m after three 20\,min temperature cycles. 
The electric field and transient simulations have been validated according to this change in size and show a negligible effect on the $\etaup$-function.

\section{Conclusions}
Enhanced Lateral Drift sensors feature a dedicated charge sharing mechanism in order to improve the resolution of the impact position of ionising particles without using a magnetic field or sensor tilt.
A carefully tuned buried implant structure creates a non-homogeneous electric field in the lateral direction in the bulk of the sensor changing the drift path of the charge carriers.
This enables charge sharing and hence improved position resolution in thin sensors. 

Two types of sensors, n-ELAD and p-ELAD sensors, have been simulated including the buried implant structure. 
TCAD simulations for the 2D and 3D geometries show congruent results of the electric field.
The transient simulations revealed the dependence of the $\etaup$-function, and hence of the spatial resolution, on the buried implant concentration and readout implant size.
A close to linear $\etaup$-function is achieved for $n\mathrm{_{di}} = 2.8 \mathrm{\cdot10^{15}\,cm^{-3}}$ at a bias voltage of $U=280$\,V for the n-ELAD sensor 
 and for $n\mathrm{_{di}} = 2.4 \mathrm{\cdot10^{15}\,cm^{-3}}$ at a bias voltage of $U=300$\,V for the p-ELAD sensor
 demonstrating the possibility to approach the theoretical optimum of charge sharing.
The electric field profiles stemming from 2D and 3D simulations will be used as input for Monte Carlo-based position resolution studies.
Additionally, the foreseen manufacturing process for ELAD prototypes has been outlined. 

{\small
\bibliographystyle{IEEEtran}
\bibliography{bibtex/refs}
}

\end{document}